\documentclass[11pt]{article}
\usepackage{geometry}                
\usepackage{graphicx,cite}
\usepackage{amsmath,amssymb, amsthm,epsfig,bm}
\usepackage{setspace,sectsty}
\usepackage{epstopdf}
\usepackage{psfrag}
\usepackage{color,soul}
\sethlcolor{yellow}
\allsectionsfont{\centering}

\begin{document}

\title{Random Walk over Basins of Attraction to Construct \\ 
Ising Energy Landscapes}
\author{Qing Zhou\thanks{Email: zhou@stat.ucla.edu}\\
Department of Statistics, University of California, Los Angeles, CA 90095, USA}
\date{}
\maketitle

\begin{abstract}

An efficient algorithm is developed to construct disconnectivity graphs 
by a random walk over basins of attraction. This algorithm can detect a large number of
local minima, find energy barriers between them, and estimate local thermal averages over each basin of attraction.
It is applied to the SK spin glass Hamiltonian where existing methods have difficulties even for a moderate number of spins.
Finite-size results are used to make predictions in the thermodynamic limit that match theoretical approximations
and recent findings on the free energy landscapes of SK spin glasses.


\end{abstract}

Disconnectivity graphs (DGs) \cite{Becker97, Wales98}, widely used for representing energy landscapes,
summarize local minima and energy barriers of an energy function into a tree.
The DG of a continuous energy surface can be constructed by computational approaches that
search for local minima and saddles based on the gradient and the Hessian matrix \cite{Wales05}.
However, such approaches cannot be applied to Ising Hamiltonians
defined on discrete spins. For example, the Hamiltonian of the SK spin glass \cite{SK75} with zero external
magnetic field, which is the focus of this paper, is
\begin{equation}\label{eq:Hs}
H(s)=-\sum_{i<j} J_{ij}s_i s_j, 
\end{equation}
where $s=(s_1,\ldots,s_N)$, $s_i \in \{\pm 1 \}$, is a vector of $N$ spins 
and $J_{ij}$ is the interaction between $s_i$ and $s_j$.
Many studies have been conducted, such as in \cite{Bray80, Tanaka80, Aspelmier04,Cavagna04, Aspelmeier06}, 
to characterize the free energy landscape of the SK spin glass by
investigating solution structures of the TAP free energy equations \cite{Thouless77}. 
See \cite{Parisi06} for a review and more references. 
These studies reply heavily on specific assumptions 
for the distribution of the disorder $J=\{J_{ij}\}$. 
On the other hand, computational approaches have been developed to construct DGs for
spin glass Hamiltonians \cite{Nemoto88, Garstecki99, Fontanari02, Hordijk03, Burda06, Seyed08, Zhou09}. 
From DGs one may extract microscopic information to characterize free energy landscapes.
In principle, these approaches can be applied given any possible distribution of the disorder,
but, unfortunately, they are feasible only for small-scale systems with less than or around 30 spins,
due to the computationally expensive nature of DG construction. 

The purpose of this paper is to develop an efficient algorithm that is able 
to construct DGs containing hundreds of local minima for spin systems with $N$ on the order of 100 and possibly larger. 
The algorithm is motivated by the broad success of the Wang-Landau (WL) algorithm \cite{Wang01a, Wang01b}  
which produces a random walk in energy space. To build a DG, we aim to generate a random walk
over the basins of attraction of local minima. 
Suppose that the Hamiltonian $H(s)$ has $K$ local minima, $v_1,\ldots,v_K$. 
The basin of attraction of $v_k$, denoted by $D_k$, is the set of
configurations which will be sent to $v_k$ by steepest descent that recursively flips the single spin giving 
the maximum decrease in $H(s)$.
If a random walk can be produced over all basins of attraction, $D_1,\ldots,D_K$, not only do we have
all the local minima but also may estimate local thermal averages over every $D_k$.
Such estimation on basins of attraction is a key to the utility of the inherent structure approach \cite{Stillinger84,Sciortino05}
and the superposition approach \cite{Strodel08}. 
Furthermore, frequent transitions between
basins must occur during the walk. We say two configurations $x$ and $y$ are neighbors,
denoted by $x\leftrightarrow y$, if they differ by only one spin. As each local move is a single-spin flip,
cross-basin moves can be used to find the barrier between two basins defined as
$B_{k\ell} = \min_{p\in \mathcal{P}_{k\ell}} \max_{s\in p}H(s)$,
where $\mathcal{P}_{k\ell}$ is the collection of all paths between $v_k$ and $v_{\ell}$ ($k\ne \ell$) 
in the configuration space.
For a spin system, 
\begin{equation}\label{eq:barrierpair}
B_{k\ell}= \min\{H(x)\vee H(y): x \in D_k, y \in D_{\ell}, x\leftrightarrow y\},
\end{equation}
where $H(x)\vee H(y) \equiv \max[H(x),H(y)]$.
Thus, keeping track of cross-basin moves, we may obtain a rough estimate of $B_{k\ell}$ which can
be refined by a ridge descent algorithm to be introduced later. With local minima and barriers detected
constructing the DG of $H(s)$ is trivial.

Since the number of minima increases exponentially with $N$ for SK spin glasses, 
we follow the practical convention to construct DGs with $K$ lowest local minima for a big $K$.
For the sake of understanding, we first describe the algorithm in the context that $K$ local minima
of $H(s)$ have already been detected. These local minima are used to partition the space into 
$K$ basins, $D_1,\ldots,D_K$, and their complement $D_0$. 
As energy of SK spin glasses is continuous, a ladder of energies,
$u_0<u_1<\cdots<u_L=\infty$, where $u_0$ is a lower bound of $H(s)$, 
are employed to partition the energy space into $L$ intervals.
Then, our goal is to generate a random walk over all (nonempty) subregions,
$D_{kj}=\{s \in D_k: H(s) \in [u_{j-1},u_j)\}$, $k=0,\ldots,K$ and $j=1,\ldots,L$,
where two indices, the basin index $k$ and the energy index $j$, are used for space partition.
The desired random walk can be implemented by
a generalized WL (GWL) algorithm \cite{Liang07,Atchade10},
where energy on a subregion $D_{kj}$ is not a constant.
Let $\theta_{kj}$ denote the (unnormalized) statistical weight of $D_{kj}$ in the Boltzmann distribution,
i.e., $\theta_{kj}\propto \sum_{D_{kj}}e^{-\beta H(s)}$, where $\beta$ is the inverse temperature.
A flat histogram over all $D_{kj}$ can be produced if the probability of visiting
$s \in D_{kj}$ is proportional to $e^{-\beta H(s)}/\theta_{kj}$. Since $\theta_{kj}$ is unknown we set $\theta_{kj}^{(1)}=1$
at the first iteration of the walk.
At iteration $t$, let $\theta^{(t)}_{kj}$ be the estimate of $\theta_{kj}$ and $x_t$ and $y$ be the configurations
before and after a randomly chosen spin is flipped.
A steepest descent operation is applied on $y$ to find its basin index, in which the energy change of a single-spin flip
can be computed efficiently by utilizing the additive structure in \eqref{eq:Hs}. If $y$ is not in the basin of any
of the $K$ minima, then $y \in D_0$. In general, if $x_t\in D_{kj}$ and $y \in D_{\ell i}$,
the Metropolis ratio from $x_t$ to $y$ is
\begin{equation}\label{eq:GWLupdate}
r(x_t\to y) = \min \left\{1,e^{\beta[H(x_t)-H(y)]}\,{\theta^{(t)}_{kj}}/{\theta^{(t)}_{\ell i}}\right\}.
\end{equation}
Each time a subregion $D_{kj}$ is visited, the weight $\theta^{(t)}_{kj}$ will be updated to 
$\theta^{(t+1)}_{kj}=\theta^{(t)}_{kj}f$ with a modification
factor $f>1$. 
Following the WL algorithm, $f$ is reduced to $\sqrt{f}$ when the flatness of the
histogram becomes acceptable (maximal fluctuation $<25\%$) 
since the last reduction of $f$. 
If the energy ladder is dense enough such that the energy in $[u_{j-1},u_j)$ is approximately
a constant, then the local density of states $\Omega_{kj} \propto \theta_{kj}  e^{\beta u_{j-1}}$, where $\Omega_{kj}$
is the number of configurations in the basin $D_k$ with energy $u_{j-1}$. In this scenario,
local thermal averages over a basin of any temperature can be obtained via estimated $\theta_{kj}$,
similar to the calculations in \cite{Wang01a,Wang01b,Zhou09}.

Suppose the random walk has been simulated for $n$ iterations.
Let $(x_t,x_{t'})$ be a pair of configurations simulated at two consecutive iterations, i.e., $|t-t'|=1$.
For any two basins $D_k$ and $D_{\ell}$ we keep track of the configuration pair
\begin{equation*}
(a,b)_{k\ell}=\arg\min_{(x_t,x_{t'})} \{H(x_t)\vee H(x_{t'}): x_t\in D_k, x_{t'} \in D_{\ell}\},
\end{equation*}
for $1\leq t,t' \leq n$.
At the last iteration, $(a,b)_{k\ell}$ is the pair that minimizes \eqref{eq:barrierpair} among
all cross-basin neighbors generated by moves between the two basins,
which provides a rough estimate of $B_{k\ell}$. A ridge descent algorithm is developed to refine the 
estimate. Let $(a_0,b_0)=(a,b)_{k\ell}$ such that $a_0 \in D_k$ and $b_0\in D_{\ell}$. For $t=1,2,\ldots$, find iteratively
\begin{eqnarray*}
a_t  & =  & \arg \min_a \{H(a): a \in \mbox{Ngb}(b_{t-1}) \cap D_k \}, \\
b_t  & =  & \arg \min_b \{H(b): b \in \mbox{Ngb}(a_{t}) \cap D_{\ell} \},
\end{eqnarray*}
until $b_{t-1}=b_t$, where $\mbox{Ngb}(s)$ is the set of all the neighbors of $s$. This iterative algorithm moves
$(a_0,b_0)$ downhill along the ridge separating the two basins. For every pair of $k$ and $\ell$, the barrier
$B_{k\ell}$ will be estimated by $H(a_t)\vee H(b_t)$ at the final iteration of the ridge descent.

Next we discuss how to identify $K$ local minima in the burn-in period of the random walk.
A collection of local minima, $V$, is dynamically accumulated using 
the same GWL update \eqref{eq:GWLupdate} in the burn-in period, but with a constant modification factor $f \equiv e$. 
Initially, $V$ is empty. In every iteration,
a local minimum is located by steepest descent starting from the proposed configuration $y$. 
Then, $V$ is updated to include the lowest $K$ minima identified so far
or all of them if there are less than $K$ minima identified.
With $\ln f \equiv 1$, $\ln \theta_{kj}^{(t)}$ simply records the
number of visits to $D_{kj}$. This makes it easy to update these weights when the local minima
in $V$ are updated. As pointed out in many previous studies, the WL update with a big
$f$ enables the walk to reach all subregions very quickly,
which is the key to detecting sufficient low-energy minima.

The proposed algorithm may be tested on the SK spin glass model \eqref{eq:Hs}, where the
$J_{ij}$ are independent Gaussian random variables with mean 0 and variance $1/N$.
Hereafter, only rescaled energy (energy per spin) will be used.
We constructed DGs for small-scale systems with $N=25$ for which exact results can be obtained
via enumeration as well as larger $N$ where enumeration is impossible. 
For each $N$ the algorithm was applied to 100 independent samples of $J=\{J_{ij}\}$ with $\beta = 1/T_c=1$.
A rough energy range of interest of this model is $[-0.8, -0.3]$. Accordingly, 
the energy space was partitioned into $L=10$ intervals with
$u_j=-0.8+j\Delta u$ for $j=0,\ldots,9$, where $\Delta u=0.1$ for $N\leq 60$ 
and $\Delta u=0.05$ for $N\geq 70$.

For $N=25$, our algorithm was applied to each sample with  a total of $1\times 10^7$ MC sweeps. 
We chose $K=500$ which turned out to be greater than the total
number of minima for all the samples, ranging from 56 to 310.
Compared to results from enumeration, the constructed DGs were highly accurate. 
Our algorithm did not miss a single minimum for
any sample. Recall that due to the use of steepest descent 
our algorithm will not produce any false minima. The average absolute energy difference 
between estimated and exact barriers was $1.4\times 10^{-7}$, which
was extremely small relative to the energy range of the model.
This demonstrates that the algorithm indeed
accurately recovered most energy barriers. In fact, 
our algorithm recovered exactly all the barriers for 99 out of the 100 samples. 
Finding barriers is a difficult job especially for discrete Hamiltonians. The result here highlights
the advantage of simulating a random walk over basins of attraction in building DGs.
The average acceptance rate for the MC moves was $\sim$30\%. 
Thus, by exploring only 9\% of all configurations 
our algorithm was able to construct DGs almost identical to those by enumeration.

\begin{figure}[t]
\centering
\includegraphics[width=6.5in, trim=0in 1in 0in 0.8in,clip]{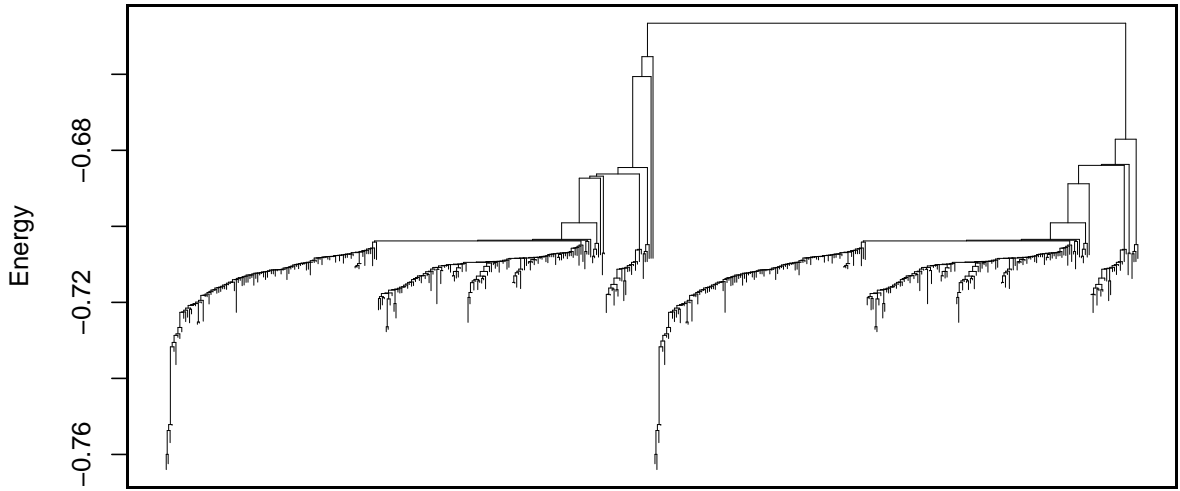}\\ 
\caption{A constructed DG for the SK spin glass with $N=100$. 
A terminal node (leaf) on the tree represents a local minimum and an internal node (branch point)
represents an energy barrier, with energy levels given by the vertical axis. The statistical error
between independent runs is very small (see text for more discussion).
\label{fig:trees}} 
\end{figure}

We applied our algorithm to $N=40,50,\ldots,100$, aiming at constructing DGs for the lowest
$K=500$ minima. Each run consisted of $5\times 10^8$ MC sweeps. 
The acceptance rate for MC sweeps was $>15\%$ for each $N$, averaging over the samples.
At the final iteration, $\ln f$ decreased below $10^{-6}$ for most of the samples
with $N \leq 60$ and was on the order of $10^{-5}$ to $10^{-3}$ for $N\geq 70$ (Table~\ref{tab:Nbig}).
These results suggest that our algorithm well explored identified basins and made
frequent transitions between them, which is sufficient for constructing DGs
although estimation of the weights $\theta_{kj}$ may not be very accurate for large $N$.
Figure~\ref{fig:trees} shows the DG constructed for a sample with $N=100$.
One sees two almost identical subtrees,
each containing a ground state and a few groups of local minima, joining at the highest detected
barrier. The identical structure between the two subtrees, due to the fact that $H(-s)=H(s)$ \eqref{eq:Hs},
gives a validation of the DG.
However, the algorithm did not recover
the energy landscape for those missing high-energy minima. 
This limitation is inevitable due to the exponential increase in the complexity
of SK spin glasses. To quantify the statistical error of a constructed DG, we applied independently our algorithm to this sample
ten times. Remarkably, all the identified local minima and at least 95\% of the estimated barriers 
were exactly identical between any two runs. Furthermore, we systematically compared the two subtrees 
of a DG to measure the accuracy of our algorithm. For all $N$,
the two subtrees of every sample contained identical sets of local minima, up to reversal of all the spins,
and substantially overlapping sets of barriers (Table~\ref{tab:Nbig}),
which demonstrates the reliability of the constructed DGs.

\begin{table}[t]
\caption{Convergence and accuracy for $N\geq 40$. 
Note: $-\log(\ln f)$ is the negative logarithm (base 10) of the median, over 100 samples, of $\ln f$ at the end of simulation;
$\eta$ is the percentage of barriers that are identical
between the two subtrees of a DG, averaging over 100 samples.
 \label{tab:Nbig}}
\centering
\vspace{0.05in}
\begin{tabular}{cccccccc} 
   \hline
   $N$ & 40 & 50 & 60 & 70 & 80 & 90 & 100 \\
   \hline
   $-\log(\ln f)$ & $>6$ & $>6$ & $>6$ & 5.4 & 4.8 & 4.2 & 3.9 \\
   $\eta$(\%) & 100 & 99.7 & 99.1 & 99.3 & 97.6 & 91.2 & 85.0 \\
\hline
\end{tabular}
\end{table}

\begin{figure}[t]
\centering
  \begin{minipage}[b]{2.5in}
   \centering
   \includegraphics[width=\linewidth,trim=0in 0.2in 0.4in 0.7in,clip]{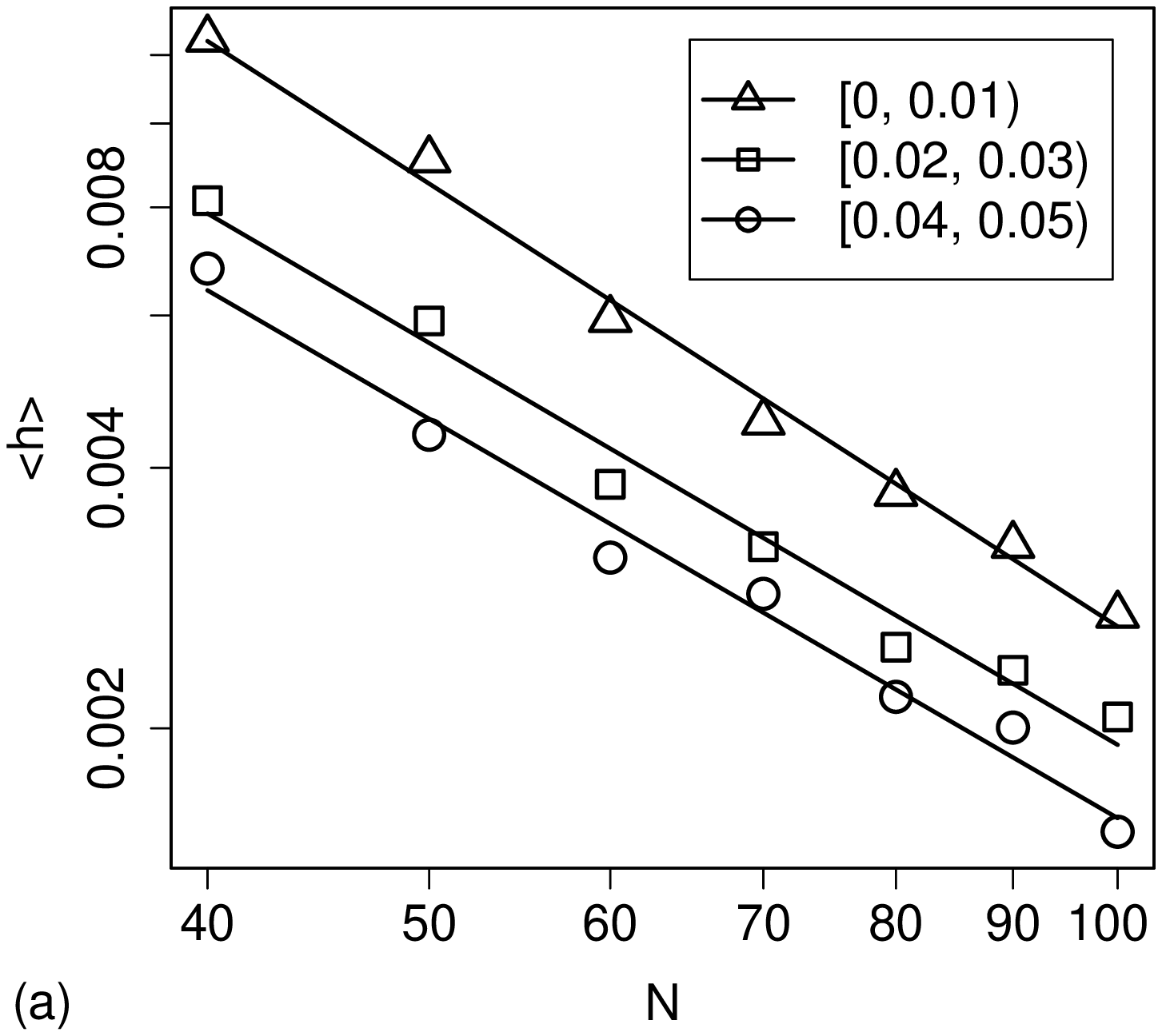} \\
\end{minipage}
\hspace{0.1in}
\begin{minipage}[b]{2.5in}
   \centering
   \includegraphics[width=\linewidth,trim=0in 0.2in 0.4in 0.7in,clip]{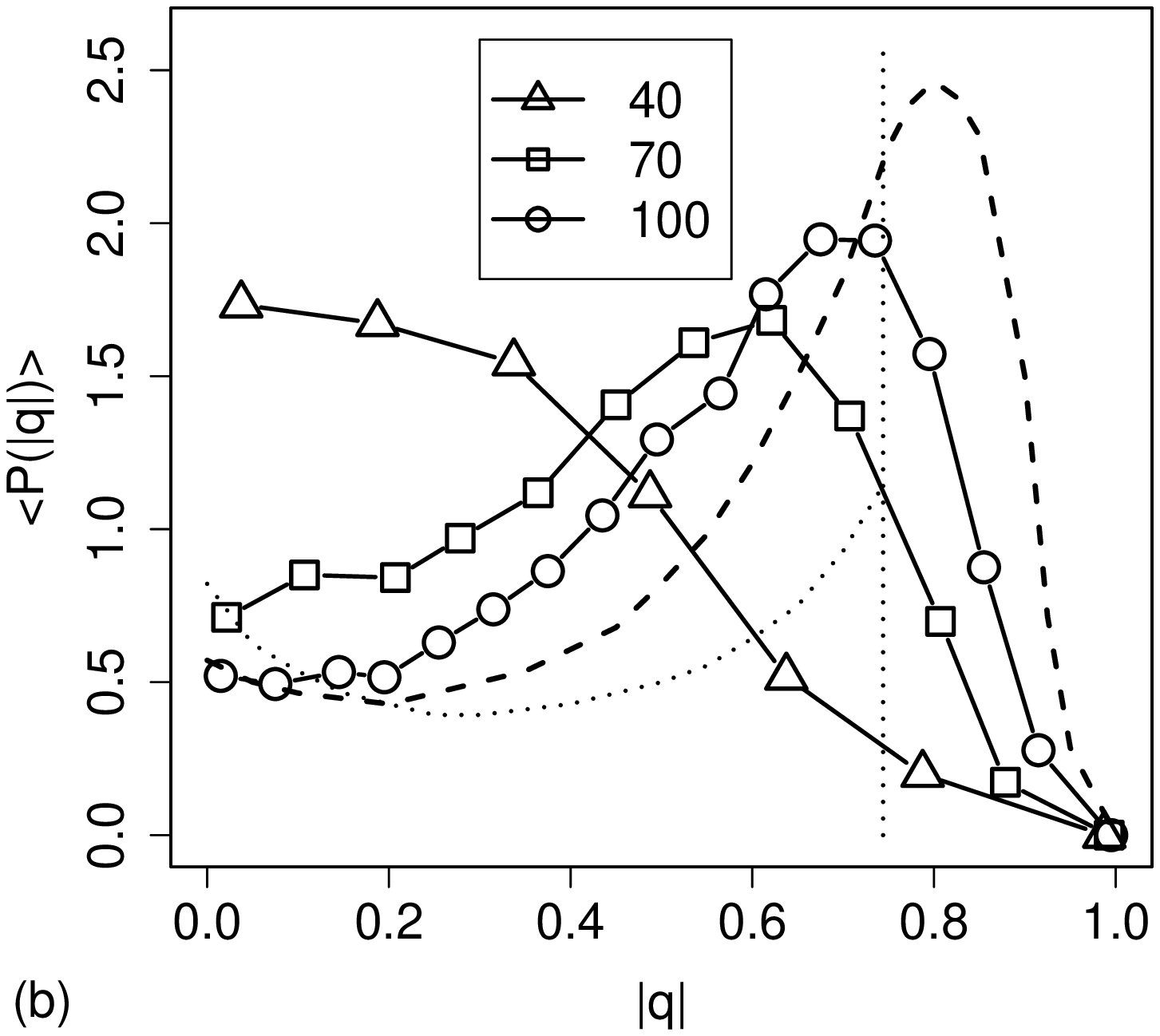} \\
\end{minipage}   
   \caption{Results for $N\geq 40$. (a) Average barrier height $\langle h \rangle$ 
  and its fitted power laws for three groups of minima with energy ranges
   given by the intervals in the legend. Log scale is used for both axes.
   (b) $\langle P(|q|)\rangle$ for $T=0.4$.
   Solid lines with symbols are results for $N=40,70$ and 100 from this study. 
   Dashed and dotted lines are, respectively, the result for $N=128$ obtained
   by Monte Carlo simulation and the result of Parisi's prediction as $N\to\infty$,
   both from \cite{Young83}.
    \label{fig:curvefit}}
\end{figure}

Define the barrier height $h$ of a minimum as the energy difference
between the minimum and its nearest barrier (its parent on the tree).
We grouped minima according to their energy $u$ 
relative to the global minimum $u^*$ and studied the relation between $\langle h\rangle$ and $N$ 
for each group, where $\langle X \rangle$ denotes the average of $X$ over samples of the disorder. 
We analyzed five groups of minima with $(u-u^*)\in [0.01(z-1),0.01z)$ for $z=1,\ldots,5$.
In each of  these energy intervals, our algorithm detected more than 1000 minima over the 100 samples
for $N=100$. A power law, $\langle h\rangle=c N^{\lambda}$, was fitted
with extremely high $R^2(>0.98)$ for each group, where $R$ is the correlation coefficient between $\ln\langle h\rangle$
and $\ln N$. The high consistency across different $N$ serves as a confirmation for the accuracy of this result.
Figure~\ref{fig:curvefit}(a) shows the fitted power laws for three groups ($z=1,3,5$),
from which we see the three lines are almost parallel to each other and 
that $\langle h \rangle$ clearly decreases with the increase of the energy of a minimum. 
The estimated $\lambda$ for $z=1,\ldots,5$ were $-1.70\pm0.06$, $-1.63\pm0.07$, $-1.54\pm0.09$,
$-1.47\pm0.08$ and $-1.53\pm0.09$, respectively. These powers were not significantly different especially
between neighboring groups. This result implies that the barrier height of a minimum vanishes as $N\to \infty$ and the rate
of decay is comparable among minima with different energy. Thus, all minima become marginally
stable as $N\to\infty$, which is consistent with the recent finding that each minimum 
and its nearby saddle on the free energy surface 
of the SK spin glass coalesce in the thermodynamic limit \cite{Aspelmier04,Cavagna04}.

A constructed DG may also provide qualitative understanding about pure states. When the temperature $T$ is low,
it is reasonable to approximate a pure state $\alpha$ by a local minimum
and determine its statistical weight by $w_{\alpha} \propto e^{-Nu_{\alpha}/T}$,
where $u_{\alpha}$ is the energy of the local minimum. 
Then, one may find the probability distribution, 
$P(q)=\sum_{\alpha,\gamma} w_{\alpha}w_{\gamma} \delta (q_{\alpha\gamma}-q)$,
for the overlap $q_{\alpha\gamma}$ between two pure states, 
$\alpha$ and $\gamma$, and its average, $\langle P(q)\rangle$, over samples of the disorder.
From the constructed DGs we approximated $\langle P(q)\rangle$ for $T=0.4$. 
Since the distribution is symmetric around $q=0$, we plot $\langle P(|q|) \rangle $ in Figure~\ref{fig:curvefit}(b).
Our approximation is compared against direct Monte Carlo simulation for $N=128$ and Parisi's prediction
as $N\to\infty$ under the same temperature \cite{Young83}. 
With the increase of $N$, the overlap distribution from our approximation becomes closer to the
expected shape and the location of the peak for $N=100$  
is in good agreement with Parisi's prediction.
This shows the utility of DGs in characterizing 
the key features (e.g., the order parameter) of spin glasses for low temperature.
However, when the temperature is high, say close to $T_c$, 
the statistical weight of the missing high-energy minima will be larger and this approximation
is likely to underestimate $P(|q|)$ for small $|q|$. 

Computational approaches that combine local optimization and Monte Carlo sampling,
such as this work,
have been developed for global optimization with applications to protein and peptide models 
\cite{Li87,Wales99,Massen07}.
These existing methods were not designed to construct DGs for Ising spin models and
are different in nature from this work. In addition, the basin-sampling approach \cite{Bogdan06}
employs the WL algorithm to construct the total energy density of states, which shares some common
features with the present work.
Although we have focused on the SK spin glass with Gaussian interactions, 
it should be noted that our algorithm is applicable
to other possible choices of the disorder $J$ and many other spin systems. 
For a continuous system, our method can be employed to
find local minima with a suitable local optimization algorithm and provide rough energy barriers which
may be refined with alternative geometry optimization methods \cite{Wales05}. 

I thank the two referees for their helpful suggestions and comments.
This work was supported by NSF grant DMS-0805491.

\singlespacing


\begin{thebibliography}{99}

\bibitem{Becker97} O.M. Becker and M. Karplus, { J. Chem. Phys.} {\bf 106}, 1495 (1997).

\bibitem{Wales98} D.J. Wales, M.A. Miller, and T.R. Walsh, Nature {\bf 394}, 758 (1998).

\bibitem{Wales05} D.J. Wales, Phil. Trans. Roy. Soc. A {\bf 363}, 357 (2005).

\bibitem{SK75} D. Sherrington and S. Kirkpatrick, 
{Phys. Rev. Lett.} {\bf 35}, 1792 (1975).

\bibitem{Bray80} A.J. Bray and M.A. Moore, {J. Phys. C: Solid St. Phys.} {\bf 13}, L469 (1980).

\bibitem{Tanaka80} F. Tanaka and S.F. Edwards, {J. Phys. F: Metal Phys.} {\bf 10}, 2769 (1980).

\bibitem{Aspelmier04} T. Aspelmeier, A.J. Bray and M.A. Moore, Phys. Rev. Lett. {\bf 92}, 087203 (2004).

\bibitem{Cavagna04} A. Cavagna, I. Giardina, and G. Parisi, Phys. Rev. Lett. {\bf 92}, 120603 (2004).

\bibitem{Aspelmeier06} T. Aspelmeier, R.A. Blythe, A.J. Bray, and M.A. Moore, 
Phys. Rev. B {\bf 74}, 184411 (2006).

\bibitem{Thouless77} D.J. Thouless, P.W. Anderson, and R.G. Palmer, Philos. Mag. {\bf 35}, 593 (1977).

\bibitem{Parisi06} G. Parisi, Proc. Natl. Acad. Sci. {\bf 103}, 7948 (2006).

\bibitem{Nemoto88} K. Nemoto, {J. Phys. A} {\bf 21}, L287 (1988).

\bibitem{Garstecki99} P. Garstecki, T.X. Hoang, and M. Cieplak, {Phys. Rev. E} {\bf 60}, 3219 (1999).

\bibitem{Fontanari02} J.F. Fontanari and P.F. Stadler, 
J. Phys. A: Math. Gen. {\bf 35}, 1509 (2002).

\bibitem{Hordijk03} W. Hordijk, J.F. Fontanari, and P.F. Stadler, { J. Phys. A} {\bf 36}, 3671 (2003). 

\bibitem{Burda06} Z. Burda, A. Krzywicki, O.C. Martin, and Z. Tabor, 
Phys. Rev. E {\bf 73}, 036110 (2006).

\bibitem{Seyed08} H. Seyed-allaei, H. Seyed-allaei, and M.R. Ejtehadi, 
Phys. Rev. E {\bf 77}, 031105 (2008).

\bibitem{Zhou09} Q. Zhou and W.H. Wong, Phys. Rev. E {\bf 79}, 051117 (2009).

\bibitem{Wang01a} F. Wang and D.P. Landau, Phys. Rev. Lett. {\bf 86}, 2050 (2001).

\bibitem{Wang01b} F. Wang and D.P. Landau, Phys. Rev. E {\bf 64}, 056101 (2001).

\bibitem{Stillinger84} F.H. Stillinger and T.A. Weber, Science {\bf 225}, 983 (1984).

\bibitem{Sciortino05} F. Sciortino, J. Stat. Mech: Theory Exp., P05015 (2005).

\bibitem{Strodel08} B. Strodel and D.J. Wales, Chem. Phys. Lett. {\bf 466}, 105 (2008).

\bibitem{Liang07} F. Liang, C. Liu, and J. Carroll, { J. Amer. Statist. Assoc.} {\bf 102}, 305 (2007).

\bibitem{Atchade10} Y.F. Atchade and J.S. Liu, {Stat. Sinica} {\bf 20}, 209 (2010).

\bibitem{Young83} A.P. Young, Phys. Rev. Lett. {\bf 51}, 1206 (1983).

\bibitem{Li87} Z. Li and Scheraga, Proc. Natl. Acad. Sci. {\bf 84}, 6611 (1987).

\bibitem{Wales99} D.J. Wales and J.P.K. Doye, J. Phys. Chem. A {\bf 101}, 5111(1997).

\bibitem{Massen07} C.P. Massen and J.P.K. Doye, Phys. Rev. E {\bf 75}, 037101 (2007).

\bibitem{Bogdan06} T.V. Bogdan, D.J. Wales and F. Calvo, J. Chem. Phys. {\bf 124}, 044102 (2006).




\end{thebibliography}
\end{document}